# FPGA-based trigger system for the Fermilab SeaQuest experiment

Shiuan-Hal Shiu[a,b,1], Jinyuan Wu[c], Randall Evan McClellan[d], Ting-Hua Chang[a], Wen-Chen Chang[a], Yen-Chu Chen[a], Ron Gilman[e], Kenichi Nakano[f], Jen-Chieh Peng[d], and Su-Yin Wang[a,c,g]

[a] *Institute of Physics, Academia Sinica,*
*128 Sec. 2, Academia Rd., Nankang, Taipei 11529, Taiwan*

[b] *Department of Physics, National Central University*
*No. 300, Jhongda Rd., Jhongli District, Taoyuan City 32001, Taiwan*

[c] *Fermi National Accelerator Laboratory*
*Kirk and Pine streets, Batavia, IL 60510-5011, USA*

[d] *Department of Physics, University of Illinois at Urbana-Champaign*
*1110 W. Green St., Urbana, IL 61801-3080, USA*

[e] *Rutgers, The State University of New Jersey*
*136 Frelinghuysen Rd., Piscataway, NJ 08854, USA*

[f] *Department of Physics, Tokyo Institute of Technology*
*2-12-1 Ookayama, Meguro-ku, Tokyo 152-8550, Japan*

[g] *Department of Physics, National Kaohsiung Normal University,*
*No.62, Shenjhong Rd., Yanchao Township, Kaohsiung County 824, Taiwan*

ABSTRACT: The SeaQuest experiment (Fermilab E906) detects pairs of energetic $\mu^+$ and $\mu^-$ produced in 120 GeV/c proton-nucleon interactions in a high rate environment. The trigger system consists of several arrays of scintillator hodoscopes and a set of field-programmable gate array (FPGA) based VMEbus modules. Signals from up to 96 channels of hodoscope are digitized by each FPGA with a 1-ns resolution using the time-to-digital convertor (TDC) firmware. The delay of the TDC output can be adjusted channel-by-channel in 1-ns steps and then re-aligned with the beam RF clock. The hit pattern on the hodoscope planes is then examined against pre-determined trigger matrices to identify candidate muon tracks. Information on the candidate tracks is sent to the 2nd-level FPGA-based track correlator to find candidate di-muon events. The design and implementation of the FPGA-based trigger system for SeaQuest experiment are presented.

## 1. Introduction

The SeaQuest experiment uses the 120-GeV/c proton beam extracted from the Fermilab Main Injector to measure muon pairs from the Drell-Yan process [1] in collisions with liquid hydrogen (p-p), deuterium (p-d), and the other nuclear targets (p-A) [2-5]. In the Drell-Yan process, quark and antiquark from colliding hadrons annihilate into a virtual photon, which then decays into a pair of oppositely charged muons. From measurement of the muon pairs, information on the quark and antiquark structure of the hadrons can be obtained. In the SeaQuest experiment, data from liquid hydrogen and liquid deuterium targets will be sensitive to the distributions of the light anti-quarks, $\bar{u}$ and $\bar{d}$, in the proton. In addition, information on the energy loss of a fast parton (quark) traversing the cold nuclear matter, and the EMC-effect for sea-quarks, can be extracted using nuclear targets.

---

[1] Corresponding author.
Tel: +886-2-2789-8944 ; E-mail : shshiu@phys.sinica.edu.tw

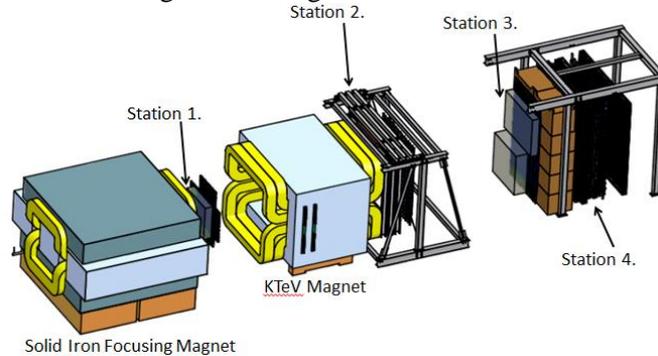

**Fig. 1.** The schematic layout of the Fermilab SeaQuest spectrometer.

The design of the SeaQuest experiment is similar to the previous Fermilab fixed-target experiments (E772, E789 and E866/NuSea reviewed in [6]) which used an 800 GeV/c proton beam. A different proton beam line of 120 GeV/c and a new spectrometer located at a different experimental hall are utilized for SeaQuest. The spectrometer layout is shown in Fig. 1. A dipole magnet labeled "Solid Iron Focusing Magnet" (FMAG), which focuses muon pairs into the detector acceptance and sweeps out low-momentum background particles, also functions as the hadron absorber and beam dump. A large open-aperture analyzing magnet labeled "KTeV Magnet" (KMAG) is for precise determination of the momenta of muon tracks. The magnetic fields of FMAG and KMAG are in the vertical (Y) direction to bend particles horizontally parallel to the X-direction. Stations 1, 2 and 3 consist of scintillation hodoscopes and multi-wire drift chambers (MWDC) for tagging and measuring the trajectories of muons. Further downstream, a large iron absorber instrumented with proportional tubes (Station 4) provides additional muon identification. One of the scintillation hodoscope arrays in SeaQuest is shown in Fig. 2. Valid Drell-Yan dimuon events will have hits on different hodoscope planes satisfying various trigger-roads (firing patterns). This is the underlying principle for the SeaQuest trigger system.

The SeaQuest beam consists of ~$10^{13}$ protons in a 5-second long slow-extraction spill once every minute. The microscopic spill structure is such that the protons are delivered in 1-ns wide "RF buckets" separated by 18.9 ns (53 MHz), known as the RF clock period. The Drell-Yan process is electromagnetic in nature and represents only a tiny fraction of the total interaction cross section. The task of the trigger system is to tag events containing a pair of opposite-sign muons using the fast signals from the scintillation hodoscopes. The trigger system should also have sufficient timing resolution to resolve hodoscope hits from adjacent RF buckets. Scintillation hodoscopes placed at each station provide prompt signals as input for the trigger modules.

While the trigger system aims at measuring events with opposite-sign muon pairs originating from the Drell-Yan process, other opposite-sign muon pairs from the decay of charmonium state (J/ψ) are also of interest. The main background comes from random coincidence of single muons from decays of mesons (π, K, D, etc.). In order to discriminate the Drell-Yan and charmonium signals from the background, a trigger matrix obtained from Monte-Carlo simulation of the Drell-Yan and J/ψ events was implemented into the FPGAs.

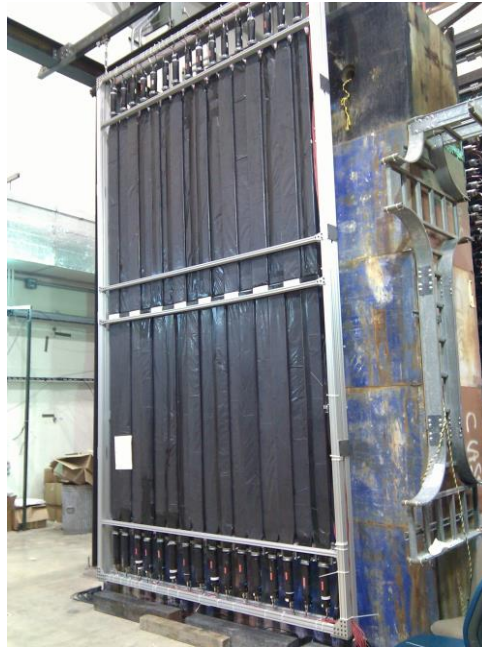

**Fig. 2.** One of the scintillation hodoscope arrays for X (horizontal)-direction tracking in the SeaQuest experiment.

In the remainder of this paper, we focus our discussion on the design of the FPGA firmware. We first introduce the overall structure of the trigger system and its basic functionality. Then we will describe the FPGA trigger module logic. The three major functions (TDC, delay adjustment pipeline, and trigger matrix) which are implemented in the trigger module logic will be discussed.

## 2. Overall structure of the trigger system

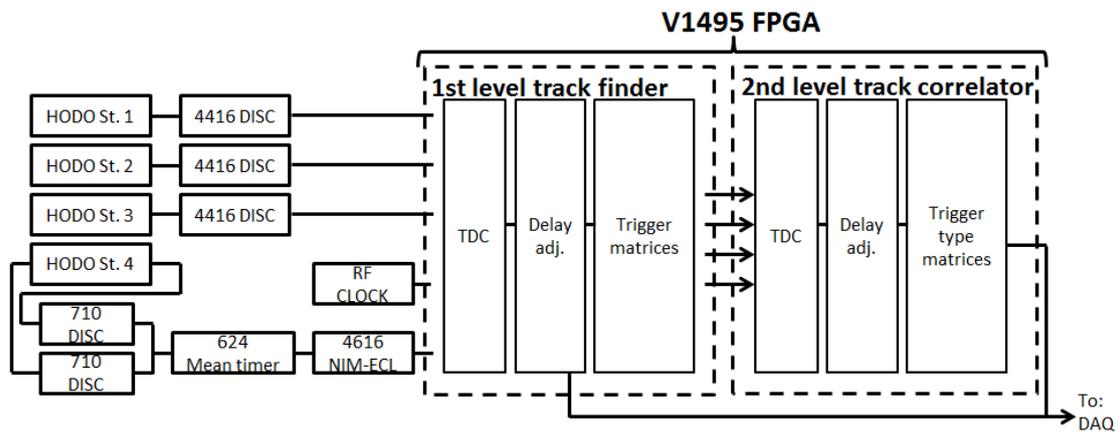

**Fig. 3.** Block diagram of the SeaQuest trigger system.

Figure 3 shows the block diagram of the SeaQuest trigger system. It consists of 4 scintillation hodoscope arrays with corresponding discriminators, the 1st level track finder, and the 2nd level track correlator. For the station 4 hodoscope, each scintillator is connected to PMTs at both ends followed by a mean timer. The specifications of the hodoscope planes are listed in table 1. The X and Y hodoscopes measure hits in the bending and non-bending planes, respectively. Each X hodoscope consists of the up (U) and down (D) parts, as shown in Fig. 2. Similarly, the Y hodoscope is divided into the left (L) and right (R) parts. A muon originating from the target forms a straight line in the non-bending Y direction measured by the Y hodoscopes. The hits on the X hodoscopes corresponds to a non-straight track related to muon's momentum. The track finding relies mainly on the hit pattern of the X hodoscopes, while the expected straight track on the Y hodoscopes could provide further track identification.

| Hodoscope | Paddle width (cm) | Paddle length (cm) | # of paddles |
|---|---|---|---|
| 1X(U/D) | 7.32 | 69.9 | 23 |
| 1Y(L/R) | 7.32 | 78.7 | 20 |
| 2X(U/D) | 13 | 120.7 | 16 |
| 2Y(L/R) | 13 | 101.6 | 19 |
| 3X(U/D) | 13.34 | 162.56 | 16 |
| 3Y(L/R) | 13.34 | 162.56 | 16 |
| 4Y(L/R) | 23.50 | 185.5 | 16 |
| 4X(U/D) | 19.69 | 182.88 | 16 |

Tab. 1. Specifications of the SeaQuest hodoscope arrays. Each hodoscope plane is divided into up/down (U/D) or left/right (L/R) halves. The number of paddles refers to each half of the hodoscope plane.

The signals from hodoscope arrays and the RF clock are sent to the FPGA as inputs for the 1st level track finder. The track-finder logic in the SeaQuest trigger system is based on the hit pattern of hodoscopes placed throughout the spectrometer. In order to analyze the hit pattern of potential dimuon events, signals from each channel should be aligned to the appropriate RF clock, associated with the collision time. To achieve this functionality, we utilize the FPGA internal TDC function with 1-ns resolution to sample the input signal and store the time of each hit with an appropriate offset for alignment with the RF clock. The precision of the channel-by-channel offset adjustment is 1 ns. A user-defined time window selects the in-time hits taking into account the variation of signal propagation time due to the finite length of each scintillator.

Compared with the trigger system in the previous Fermilab E866 experiment [7], the new system contains fewer components but significantly greater flexibility and reliability. All the modules of trigger logic, fan in/out and trigger matrix memory, and cables for delay adjustment in E866 are now integrated into the functionality of the FPGAs in SeaQuest.

The trigger logic electronics are based on the 6U VMEbus module (CAEN V1495) [8] which contains a field-programmable gate array (FPGA) with 20,060 logic elements (Altera EP1C20F400C6) [9]. The FPGA receives up to 96 channels of inputs from the hodoscopes and digitizes the leading edge time at 1-ns (LSB) resolution using time-to-digital converter (TDC) blocks in the firmware [10-12]. A photo of the V1495 module is shown in Fig. 4.

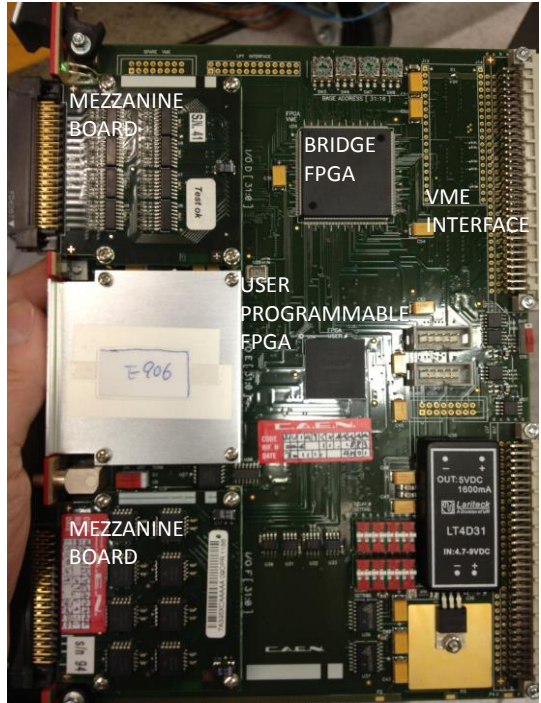

**Fig. 4.** CAEN V1495 6U VMEbus module.

The trigger system is composed of 5 individual V1495 modules. Each module, installed with input/output mezzanine boards, can provide 96 input channels and 64 output channels. Four modules are used as the 1st level track finders, which search for muon tracks in upper Y-direction hodoscopes, upper X-direction hodoscopes, lower Y's, and lower X's, respectively. For the X-direction hodoscopes, each V1495 module serves 71 inputs. For the Y-direction hodoscopes, one V1495 serves 70 inputs, the other one serves 72 inputs. The remaining module works as the 2nd level track correlator, which receives track information from the track finders and forms the final trigger decision. Based on the charge-sign and the transverse momentum ($p_T$) of the track candidates, the 2nd level track correlator will either accept or reject the events according to some predetermined firing patterns. For example, if the track correlator receives two opposite-sign muon tracks and the total $p_T$ is larger than 4.0 GeV/c, the track correlator will output a signal for a non-prescaled trigger. We also collect prescaled single-muon events for background study. During SeaQuest's Run I, we had 2 non-prescaled triggers and 4 prescaled triggers. On average, it takes ~440 ns for processing the 1st level track finder and ~330 ns for the track correlator. The specific processing time depends on the number of pipeline steps in the trigger matrices and the internal TDC delay. The ~770 ns trigger decision time is well within the 2048-ns limit set by the buffer size of the TDC modules for the drift chamber readout.

## 3. FPGA trigger module logic

The major functions of the FPGA are shown in figure 5: the TDC unit, the delay adjustment pipeline, and the trigger matrix. The 1st level track finders and the 2nd level track correlator share a common design of the TDC unit and delay adjustment pipeline. The function of the Time-to-Digital converter (TDC) is to digitize leading-edge time of the 96-channel input signals. The 4-phase sampling units are driven by a 250 MHz clock generated by the PLL (phase lock loop) function with a 40 MHz clock source. This 4-phase sampling unit can provide a 1-ns resolution TDC with a 16-ns two-hit resolution. The encoded timing information is sent to a digitizing and retiming block which synchronizes with a 62.5 MHz clock.

The digitized timing information is sent to a RAM-based pipeline structure which synchronizes the input timing with respect to the beam RF signal in 1-ns steps. Including this function in the FPGA allows all input channels to have varying signal delays due to different cable lengths and minor differences in electronics settings, and also provides a 16-ns jittering acceptance region. The RAM-based pipeline structure has a total record depth of 2048 ns. At this stage, the zero-suppressed TDC output is provided. Finally, the re-aligned signals are sent to the trigger matrix to examine the firing patterns of the hodoscopes and apply the trigger matrices.

Both the 1st and 2nd level FPGAs use the same design of these major functions, but with different contents of the trigger matrices. The VHDL (VHSIC Hardware Description Language) codes of the trigger matrices are generated automatically using a program which converts the required trigger roads into the hardware logic description. This automation reduces both the work load and potential errors during the commissioning phase of the experiment when trigger conditions are changed frequently. The details of the trigger system are presented in the following sections.

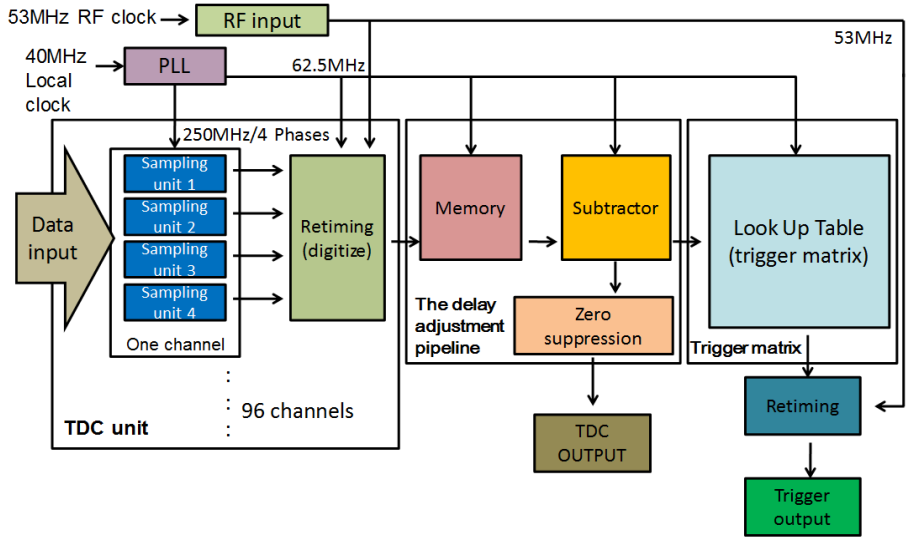

**Fig. 5.** Block diagram of the major functions in FPGA.

## 4. The TDC block

The core unit of the TDC function adopts the multi-phase sampling scheme as shown in Fig. 6. After an input buffer logic element, 4 phases (C0, C90, C180, and C270) of the 250 MHz clock are used to achieve the 1-ns timing resolution for the TDC. The four registers are connected to four internal clocks, each with a 90-degree phase difference. The 0- and 90-degree clocks are generated in a phase-lock-loop (PLL) clock synthesizer, and their inversions are used for the 180- and 270-degree clocks. The sampling interval is 1 ns even through each register operates at 250 MHz, rather than 1 GHz.

A transfer to the 0-degree clock domain occurs in the second and third stages of the pipeline. Depending on arrival time, the transitions of the input logic levels are recorded at different locations within the four registers. A detailed description of the function can be found in [13].

Due to the four clock phases, the sampling registers must be brought to the same clock domain via the clock domain-changing stage, shown in Fig. 6. Note that three registers use the c0 clock while the last one uses the c90 clock. Should the clock used in the last register be c0, the transition time between the sampling registers driven by c270 and the clock domain-changing registers would be 1 ns, which is too tight for successful firmware compilation. The transition time between the c270 and the c90 clocks is 2 ns, which is much easier to satisfy. In order to prevent ultra-short pulses caused by input circuit ringing from being digitized, the design incorporated the function of transition edge regulation. As shown in Fig. 6, the bit pattern on QD to Q3 is used by a look-up table in an FPGA logic element to determine if a sampling point is at the edge of a well-established pulse. The sampling position of the input signal edge represents the arrival time and is encoded as the lowest two bits, T0 and T1, together with a data valid (DV) bit. The higher 2 bits, labeled TS, are generated with a coarse time counter. The coarse time, fine time and data valid signals are conveyed to the later stages for further operation.

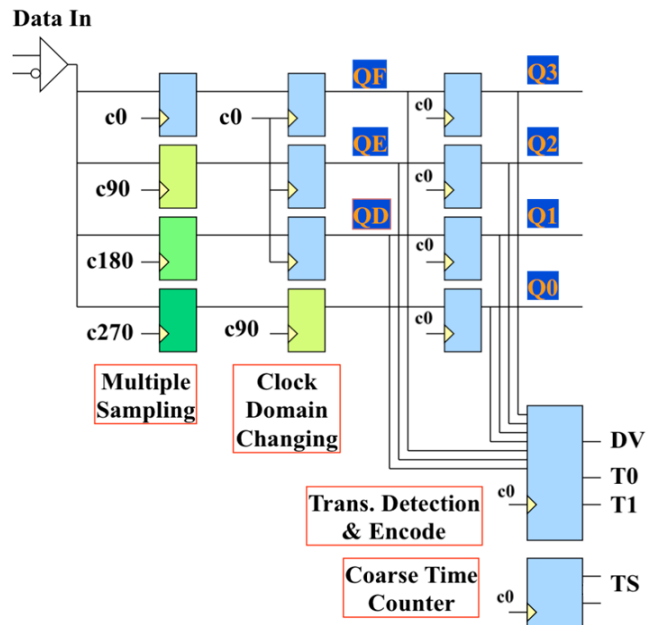

**Fig. 6.** Structures of TDC unit in the SeaQuest trigger FPGA

To ensure equal propagation delay from input buffer to the multi-sampling registers and the clock domain-transfer registers, the input cell and the registers for each channel should be placed at suitable locations in the FPGA. A Fortran program is written to accomplish this task. The output of the program is included in the assignment file for compiling the FPGA design software. Placement of other logic elements is more flexible and can be automatically performed by the FPGA compiler.

To implement 32 channels driven by a fast clock into the FPGA is straightforward. The compiler has enough resource to place the gate element and routing. But it becomes a challenge when a large number of channels are required. In developing the firmware for 96 channels, the power capability of the module prevented us from using high clock frequency in a small portion of the FPGA. We carefully designed each block with smallest possible silicon area and the clock of lowest possible frequency.

The 1-ns timing resolution for the trigger-module TDC allows us to monitor and correct for any timing shift for each scintillator. Such timing shift could affect the trigger efficiency if undetected. Moreover, the trigger-module TDC provides a cross check for another TDC system with 2.5-ns timing resolution, which was used for the hodoscope readout during part of the data taking.

## 5. The delay adjustment pipeline structure

The timing data from the TDC are sent to RAM blocks used as pipelines and also for event storage. Figure 7 shows the pipeline structure. The input delay in each channel is adjustable individually in 1-ns step. Each bin in the pipeline corresponds to a 16-ns time interval and each memory location contains 4 bits for hit time with 1-ns (LSB) resolution plus 1 data-valid (DV) bit. In each channel, a relative delay value of 0-255 ns is stored in an 8-bit register. The lowest 4 bits of this register and the 4 bits from the TDC output are summed and written into the pipeline. If the carry from the sum is 1, the data will be delayed by one clock cycle before being written into the pipeline. The highest 4 bits of the register and the pipeline pointer counter are summed as the pipeline writing address. The channel-by-channel writing operations on a pipeline memory block serving 4 channels are clocked at 250 MHz while the reading operations are parallel for the 4 channels and clocked at 62.5 MHz. In this way, individual channel delays are compensated at the output port of the pipeline.

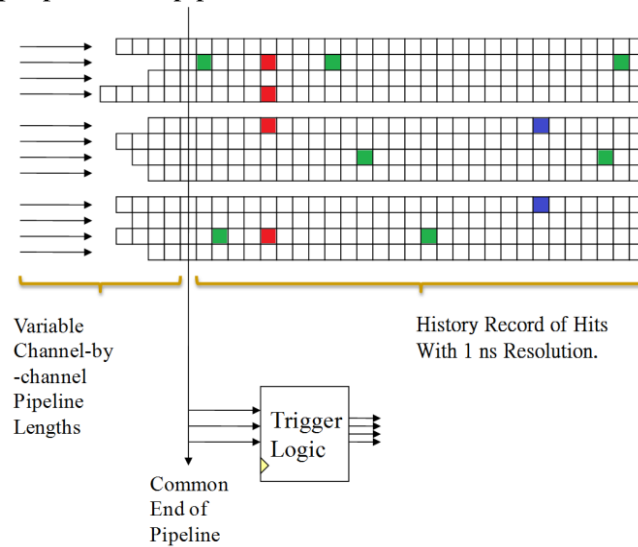

**Fig. 7.** The pipeline structure inside the FPGA. The in-time events are indicated by the red and blue blocks. The singly fired paddles are indicated by green blocks.

The pipeline is also used for event storage. When a global trigger is received, the pipeline stops; a history record of 16 time slots (TS) for all 96 channels i.e., 96*16 = 1536 words will be copied from the pipeline to the VME interface buffer at 500 Mb/s, which takes 24.576 $\mu$s. The empty time slots are suppressed during the copying process. The buffer capacity is 256 hits but the amount of data read out in each event is user defined. The copying sequence loops over the hits of channels 0-95, starting with latest time slot. Therefore, if there are more than 256 hits within the 16 time slots (which is unlikely), only the latest hits will be readout. Most of the time there are less than 256 hits and the unfilled words in the buffer will be marked as the end of block. The FPGA can be used as a zero-suppressed TDC simply by disabling the trigger matrix.

## 6. Trigger matrices

The trigger matrix is composed of logic operators like "AND", "OR", and the logic elements of hodoscope hits. The hit patterns for events of interest are determined from Monte-Carlo simulation and converted into the trigger matrix uploaded into the FPGA. The FPGA is a powerful and flexible tool to implement and adjust the desired trigger condition without changing the hardware.

We use a FORTRAN based Monte Carlo simulation code (Fast MC) to simulate all possible hit patterns of hodoscopes for events of interest. Fast MC contains information on the detector geometry, magnet configuration and field strength, and event generators for Drell-Yan, J/ψ, and charmed-hadron production. J/ψ events provide a valuable calibration tool and are of physics interest in their own right. The Fast MC simulates the momentum and position of muons at various locations in the spectrometer and treats the magnetic field and multiple scattering effects in a simplified fashion. Another simulation code (Full MC) based on Geant-4 [14] is also available. In contrast to Fast MC, the Full MC provides a more precise estimation of effects due to magnetic fields and multiple scattering. A comparison between the output of these two MC codes shows that the precision of the fast MC is sufficient for providing the inputs to the trigger matrices.

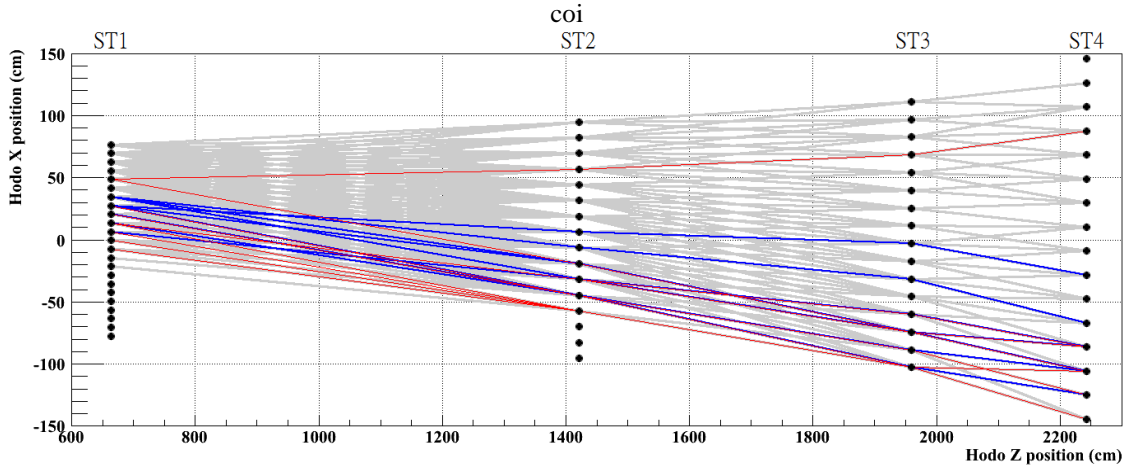

**Fig. 8.** Visualization of the hit patterns of positive muon on the X-hodoscopes. The black points represent scintillator paddles viewed from the top. The red line shows the 10 most-frequently hit patterns, and the blue lines show the next 10. Other patterns are shown in gray.

The invariant mass of Drell-Yan event is roughly proportional to the scalar sum of the transverse momenta of $\mu^+$ and $\mu^-$. A high-mass Drell-Yan event would likely lead to muons with large transverse momenta. Since the muon pairs from the meson decay background mostly have low invariant mass, they could be effectively rejected by requiring the hodoscope firing pattern to be consistent with tracks of large transverse momentum.

After identifying all potentially interesting hit patterns from the Monte Carlo study, we convert the information into a trigger matrix which is implemented in the 1st level track finder FPGA. The 1st level X-direction track finder sorts all valid patterns into transverse momentum bins. Although the Y-direction track-finders could send $p_Y$-binned track information to the 2nd level, we did not utilize the y-hodoscope information in the trigger decision during Run 1. The $p_Z$ resolution of the trigger is limited by the width of the hodoscope paddles. Any contribution to the dimuon mass from the longitudinal momenta is inaccessible to the trigger due to the relatively wide hodoscope elements.

The trigger matrix implemented in the 2nd level track correlator FPGA is designed to identify muon pairs using the $p_X$-binned tracks found in the 1st level track finder. Due the magnetic fields and the geometry of the detectors, only pairs with $\mu^+$ having positive $p_X$ and $\mu^-$ having negative $p_X$ will be accepted. Therefore, each $p_X$-bin is automatically assigned a charge. The main physics trigger requires an opposite-sign pair of tracks, with one track in the top half of the spectrometer and one track in the bottom half, satisfying some criterion on $|p_{x1}| + |p_{x2}|$. This criterion is tuned to remove the low-mass 'dimuons', which are dominated by random coincidence of low-$p_X$ single tracks. Additional outputs of the 2nd level trigger can be defined, applying various logic conditions to the 1st level outputs. Such triggers, e.g. a single-muon trigger, can be useful for detector calibration and diagnostics. Offline trigger software is capable of identifying "hot roads" in the real data, as visualized in Fig. 8. Since many hot roads are populated by background events, we disable them in the 1st level look-up table to reject backgrounds.

Each of the scintillator hodoscope detectors is assigned an ID in the trigger matrices. Based on the hit patterns of events of interest generated by the Monte Carlo code, the corresponding coincidence logic is generated. The 4-out-of-4 coincidence logic is of the form A2&B1&C1&D2. The "A2","B1","C1", and "D2" are the detector IDs; standing for four different stations and different detectors. In order to account for detector inefficiency, various 3-out-of-4 logic configurations can also be implemented. For example, the 3-out-of-4 coincidence logics are like: A2&B1&C1, A2&B1&D2, A2&C1&D2, B1&C1&D2.

When combining the 4-out-of-4 and 3-out-of-4 coincidence logic configurations in various settings of the trigger, the number of required coincidence logic elements could go up to several thousand. The compiler cannot automatically place the logic elements without generating critical timing issues. The timing issues are resolved by setting the coincidence logic output to a pipeline structure. All the pipeline logic is driven by a 250 MHz (4 ns) clock. Relative to the Fermilab main injector RF frequency of 53 MHz (18.9 ns), the trigger system is dead-time free. A diagram of the trigger matrices pipeline is shown in Fig. 9.

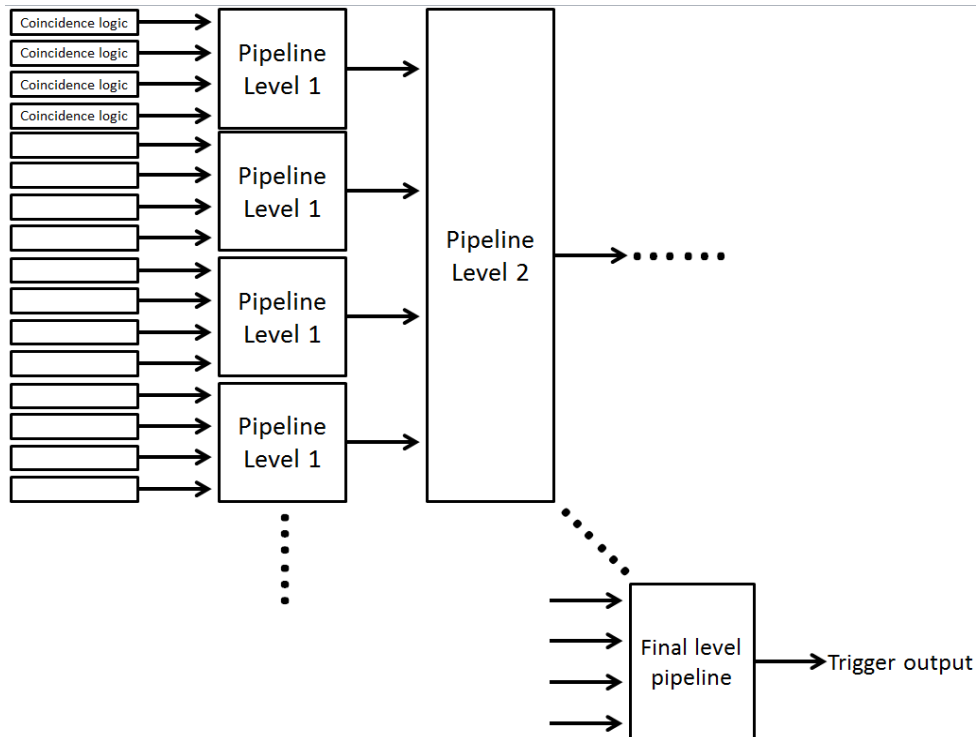

**Fig. 9.** The schematic diagram of the trigger matrix pipeline structure.

In the trigger matrices, each gate element takes four inputs, and sends the output to the next pipeline level. In the first pipeline level, the re-aligned input pattern is compared with the trigger matrices. If the input pattern satisfies any of the matrix patterns, a 'fired' result is sent to the next pipeline step. These results are OR'ed together in the subsequent pipeline steps, until finally the output bit word is issued. For the 1st level, the output is an up-to 32-bit word sent to the 2nd Level for every beam-clock cycle. Each bit represents one $p_X$-bin for 1st level tracks. For the 2nd level, the output is an up-to 5-bit word sent to the DAQ. Each bit represents a different trigger type (e.g. dimuon, single-muon). In both levels, each output bit is fired if any of the corresponding trigger matrix conditions is met. In the case that multiple conditions are satisfied for one output bit in a single beam RF clock cycle, the output bit is still fired. The compiling report shows that the pre-processing (including TDC, delay adjustment and hit realignment) takes about 8000 logic elements or 40% of the FPGA (20060 logic elements total) and the remaining resources are available for the trigger matrix.

## 7. Summary

We successfully developed the FPGA modules in the trigger system of the SeaQuest experiment with the core components of the internal TDC, delay adjustment, and trigger matrices. The internal TDC and delay adjustment enable a fast alignment of signals from each hodoscope channel with respect to the RF clock of the incident beam. The multi-phase sampling scheme utilizes the 250MHz clock driven device to implement 1-ns timing resolution and the timing critical signal paths are well controlled. Based on the delay adjustment pipeline structure, this system also provides a real-time timing adjustment function for the trigger system. The mechanism of trigger matrices provides a convenient way of implementing flexible and complicated triggers. The overall system has been proven to be reliable and flexible. During the commissioning run of SeaQuest, the performance of the SeaQuest trigger system meets the design requirements. The details of the performance will be discussed in a forthcoming paper.

## Acknowledgments

We would like to acknowledge the Fermilab E906/SeaQuest Collaboration for helpful discussions. This work was supported in part by Fermi Research Alliance, LLC under Contract No. DE-AC02-07CH11359 with the United States Department of Energy.

## References

[1] S.D. Drell and T.M. Yan, Phys. Rev. Lett. 25, 316 (1970).

[2] E. A. Hawker et al. (FNAL E866/NuSea Collaboration), Phys. Rev. Lett. 80, 3715 (1998).

[3] J.-C. Peng et al. (FNAL E866/NuSea Collaboration), Phys. Rev. D 58, 092004 (1998).

[4] R. S. Towell et al. (FNAL E866/NuSea Collaboration), Phys. Rev. D 64, 052002 (2001).

[5] Fermilab E906 proposal, Spokesperson: D. Geesaman and P. Reimer. (2006).

[6] P. L. McGaughey, J. M. Moss, and J. C. Peng, Annu. Rev. Nucl. Part. Sci. 49, (1999).

[7] C. A. Cagliardi, E. A. Hawker, R. E. Tribble, D. Koetke, P. M. Nord, P. L. McGaughey, and C. N. Brown, Nucl. Instr. Meth. A 418, (1998).

[8] C.A.E.N. V495 General Purpose VME Board., http://www.caen.it/csite/CaenProd.jsp?idmod=484&parent=11.

[9] Altera Corporation, "Cyclone Device Handbook", (2008).

[10] J. Wu, S. Hansen & Z. Shi, "ADC and TDC implemented using FPGA", Nuclear Science Symposium Conference Record, 2007 IEEE, Oct. 26 2007-Nov. 3 2007 281-286.

[11] J. Wu, "Several Key Issues on Implementing Delay Line Based TDCs Using FPGAs," IEEE Transactions on Nuclear Science, 57, 1543 (2010).

[12] H. Sadrozinski & J. Wu, "Applications of Field-Programmable Gate Arrays in Scientific Research", Taylor & Francis, December 2010.

[13] J. Wu & S. Shiu, "Field-Programmable Gate Array (FPGA) Firmware for the Fermilab E906 (SeaQuest) Trigger", Real Time Conference (RT), 2012.

[14] S. Agostinelli et al., Nucl. Instr. Meth. A 506, 250 (2003).